\documentclass[11pt,english,american,onecolumn,nofonttune]{IEEEtran}
\usepackage[T1]{fontenc}
\usepackage[latin9]{inputenc}
\usepackage{geometry}
\usepackage{color}
\usepackage{amsmath}
\usepackage{amssymb}
\usepackage{esint}

\makeatletter
\usepackage{cite}

\newcommand{\e}{\mbox{e}}

\newcommand{\fmbox}[1]{\fbox{\vbox{\vskip4.5pt\hbox{\hskip6pt{$\displaystyle #1$}\hskip6pt}\vskip3pt}}}

\let\newcommand=\providecommand

\newcommand{\sba}[2]{\raisebox{-1mm}{\hbox to 0pt{\hspace*{-4mm}\resizebox*{#1mm}{#2mm}{\includegraphics{barra.epsi}}\hss}}}

\makeatother

\usepackage{babel}
\begin{document}
\title{An Alternative Derivation of the Gaussian Noise model}
\author{Alberto Bononi and Paolo Serena \\ Università degli studi di Parma,
Parma, Italy \\ {14 Aug. 2022} }
\maketitle
\begin{abstract}
By extending the results in \emph{Bononi 2012}, we provide here a
complete alternative derivation of Turin's Gaussian Noise (GN) model
for dual-polarization dispersion uncompensated coherent optical links.
This paper contains the lecture notes used by the authors first on
July 19, 2012, and then again on june 6, 2016 at the University of
Parma.
\end{abstract}

\section{Introduction}

Goal of this paper is to provide an alternative derivation of the
results appearing in P. Johannisson's et al. \cite{Pontus} on (a
generalization of) the well-known frequency-domain GN model introduced
by Turin's group in \cite{carena_JLT}, i.e., a closed-form formula
for the power spectral density (PSD) of the nonlinear interference
(NLI) at the output of a dispersion uncompensated (DU) coherent system.
The main proof was already reported in \cite{arxiv}, and we here
make it more clear by providing more detail. The novelty here is that
we also show that the removal of the extra ``phase term'' appearing
in the output PSD formula in \cite{Pontus} is due to a change of
variable that tracks the average nonlinear phase and yields exactly
Turin's GN formula \cite{carena_JLT}.

Ref. \cite{Pontus} first proved that the GN model is indeed a first-order
regular perturbation (RP1) model fed by independent Gaussian spectral
lines. Given the equivalence of RP1 and the Volterra series model
proved in \cite{RP}, these results also explain the coincidence of
the Volterra series and the GN model approaches to the study of DU
coherent system.

\section{Frequency Domain NLI RP1 Solution}

We start from the dual-polarization (DP) single-channel first-order
Regular Perturbation (RP1) solution of the Manakov-based dispersion-managed
nonlinear Schroedinger equation (DMNLSE) \cite[Appendix 2]{OE2012}:

\begin{equation}
\tilde{\boldsymbol{U}}(L,f)=\tilde{\boldsymbol{U}}(0,f)+\tilde{\boldsymbol{U}}_{p}(L,f)\label{eq:RP1-1}
\end{equation}
where $L$ is the total link length, and the NLI \emph{p}erturbation
field is 
\begin{equation}
\tilde{\boldsymbol{U}}_{p}(L,f)=-jP_{0}\iint_{-\infty}^{\infty}\mathcal{K}(f_{1}f_{2})\tilde{\boldsymbol{U}}(0,f+f_{1})\tilde{\boldsymbol{U}}^{\dagger}(0,f+f_{1}+f_{2})\tilde{\boldsymbol{U}}(0,f+f_{2})\mbox{d} f_{1}\mbox{d} f_{2}\label{eq:RP1single}
\end{equation}
where:

i) boldface fields are 2x1 vectors containing the Fourier transforms
(denoted by a tilde) of the X and Y polarizations in the transmitter
polarization frame of reference; a dagger stands for transposition
and conjugation; and the DP field power is normalized to an arbitrary
reference power $P_{0}$ (which in \cite{Pontus} is chosen as the
\emph{per-polarization} average power);

ii) the un-normalized scalar frequency kernel is defined as (Cfr.
\cite{Pontus} eq. (45) and \cite{OE2012} eq. (25)): 
\begin{equation}
\mathcal{K}(F)\triangleq\int_{0}^{L}\gamma'(s)\mathcal{G}(s)\e^{-jC(s)(2\pi)^{2}F}ds\label{eq:DMkernel}
\end{equation}
where $F=f_{1}f_{2}$ is the product of two frequencies, $\gamma'=\frac{8}{9}\gamma$
with $\gamma$ the fiber nonlinear coefficient, $\mathcal{G}(s)$
the line power gain from $z=0$ to $z=s$, and $C(s)=-\int_{0}^{s}\beta_{2}(z)dz$
is the cumulated dispersion in the transmission fibers (with dispersion
coefficient $\beta_{2}$) up to coordinate $s$. We do not normalize
here the frequency axis, as done instead in \cite{Pontus}. In \cite{OE2012}
we use the normalized kernel
\[
\tilde{\eta}(F)=\frac{\mathcal{K}(F)}{\mathcal{K}(0)}
\]
which then at $F=0$ equals 1. The nonlinear phase referred to power
$P_{0}$ is 
\[
\Phi_{NL}=P_{0}\mathcal{K}(0).
\]

Hence by multiplying and dividing by $\mathcal{K}(0)$ we can recast
(\ref{eq:RP1single}) as
\begin{equation}
\tilde{\boldsymbol{U}}_{p}(L,f)=-j\Phi_{NL}\iint_{-\infty}^{\infty}\tilde{\eta}(f_{1}f_{2})\tilde{\boldsymbol{U}}(0,f+f_{1})\tilde{\boldsymbol{U}}^{\dagger}(0,f+f_{1}+f_{2})\tilde{\boldsymbol{U}}(0,f+f_{2})\mbox{d} f_{1}\mbox{d} f_{2}\label{eq:RP1single-1}
\end{equation}
 From (\ref{eq:RP1single-1}), the X component of the RP1 solution
writes explicitly as 
\begin{align}
\frac{\tilde{U}_{x,p}(L,f)}{-j\Phi_{NL}}= & \iint_{-\infty}^{\infty}\tilde{\eta}(f_{1}f_{2})\tilde{U}_{x}(0,f+f_{1})\tilde{U}_{x}^{*}(0,f+f_{1}+f_{2})\tilde{U}_{x}(0,f+f_{2})\mbox{d} f_{1}\mbox{d} f_{2}+\nonumber \\
 & \,\iint_{-\infty}^{\infty}\tilde{\eta}(f_{1}f_{2})\tilde{U}_{x}(0,f+f_{1})\tilde{U}_{y}^{*}(0,f+f_{1}+f_{2})\tilde{U}_{y}(0,f+f_{2})\mbox{d} f_{1}\mbox{d} f_{2}\label{eq:field_x}
\end{align}
where the first line gives the self-phase modulation (SPM) of X on
X, while the second line gives the intra-channel cross-polarization
modulation (I-XPolM) of Y on X. A perfectly dual expression for component
Y is obtained by exchanging the indices $x$ and $y$.

\section{Gaussian Assumption and Johannisson's Result}

In \cite{Pontus},\cite{carena_JLT} the key assumption is that the
input fields are the sum of independent spectral lines:
\begin{align*}
\tilde{U}_{x}(0,f) & =\sqrt{f_{0}}\sum_{k=-\infty}^{\infty}\xi_{k}\sqrt{\hat{G}_{x}(kf_{0})}\delta(f-kf_{0})\\
\tilde{U}_{y}(0,f) & =\sqrt{f_{0}}\sum_{k=-\infty}^{\infty}\zeta_{k}\sqrt{\hat{G}_{y}(kf_{0})}\delta(f-kf_{0})
\end{align*}
with $\xi_{k}$ and $\zeta_{k}$ independent identically distributed
standard (i.e. zero-mean unit variance) circular complex Gaussian
random variables (RV). Such signals do have a \emph{per-polarization}
power spectral density $\hat{G}_{x/y}(f)$ (normalized to $P_{0}$)
in the limit $f_{0}\to0$ \cite{carena_JLT}. After long statistical
averaging calculations, the authors get the power spectral density
of the $\tilde{U}_{x,p}(L,f)$ RV as per eq. (89) of reference \cite{Pontus}.
Note that our PSD $\hat{G}(f)$ is normalized such that $G(f)\equiv P_{0}\hat{G}(f)$,
where $G$ is the un-normalized PSD per polarization. Also, $P_{x}=P_{0}\int_{-\infty}^{\infty}\hat{G}_{x}(f)df$
and $P_{y}=P_{0}\int_{-\infty}^{\infty}\hat{G}_{y}(f)df$. We report
in our notation the result in \cite{Pontus}:
\begin{align}
\hat{G}_{x,p}(f)= & P_{0}^{2}\{2\iint_{-\infty}^{\infty}\left|\mathcal{K}((f_{1}-f)(f_{2}-f))\right|^{2}\hat{G}_{x}(f_{1})\hat{G}_{x}(f_{2})\hat{G}_{x}(f_{1}+f_{2}-f)\mbox{d} f_{1}\mbox{d} f_{2}\nonumber \\
+ & \iint_{-\infty}^{\infty}\left|\mathcal{K}((f_{1}-f)(f_{2}-f))\right|^{2}\hat{G}_{x}(f_{1})\hat{G}_{y}(f_{2})\hat{G}_{y}(f_{1}+f_{2}-f)\mbox{d} f_{1}\mbox{d} f_{2}\nonumber \\
+ & \mathcal{K}(0)^{2}\hat{G}_{x}(f)\left(4(\int_{-\infty}^{\infty}\hat{G}_{x}(f)df)^{2}+4\int_{-\infty}^{\infty}\hat{G}_{x}(f)df\int_{-\infty}^{\infty}\hat{G}_{y}(f)df+(\int_{-\infty}^{\infty}\hat{G}_{y}(f)df)^{2}\right)\}\label{eq:G_xp}
\end{align}
and a dual expression for Y is obtained by swapping $x\leftrightarrow y$.
Recall that $\hat{G}_{x,p}(f)$ is the NLI PSD, normalized by $P_{0}$. 

An equivalent form of (\ref{eq:G_xp}) using the normalized kernel
is the following
\begin{align}
\hat{G}_{x,p}(f)= & \Phi_{NL}^{2}\{2\iint_{-\infty}^{\infty}\left|\tilde{\eta}(f_{1}f_{2})\right|^{2}\hat{G}_{x}(f+f_{1})\hat{G}_{x}(f+f_{2})\hat{G}_{x}(f+f_{1}+f_{2})\mbox{d} f_{1}\mbox{d} f_{2}\nonumber \\
+ & \iint_{-\infty}^{\infty}\left|\tilde{\eta}(f_{1}f_{2})\right|^{2}\hat{G}_{x}(f+f_{1})\hat{G}_{y}(f+f_{2})\hat{G}_{y}(f+f_{1}+f_{2})\mbox{d} f_{1}\mbox{d} f_{2}\nonumber \\
+ & \hat{G}_{x}(f)\left(4(\int_{-\infty}^{\infty}\hat{G}_{x}(f)df)^{2}+4\int_{-\infty}^{\infty}\hat{G}_{x}(f)df\int_{-\infty}^{\infty}\hat{G}_{y}(f)df+(\int_{-\infty}^{\infty}\hat{G}_{y}(f)df)^{2}\right)\}\label{eq:G_xp-2}
\end{align}
which better shows the formal parallel with the field equation (\ref{eq:field_x}):
the field double integral in $f_{1},f_{2}$ of the product kernel-field-field$^{*}$-field
becomes a PSD double integral in $f_{1},f_{2}$ of the product squared
kernel magnitude-PSD-PSD-PSD. 

In the next section we provide an alternative proof of (\ref{eq:G_xp-2}).

\section{The New Proof}

We here present and comment the results in \cite{arxiv}. We start
from (\ref{eq:field_x}) and make the following two assumptions regarding
the input X,Y fields $U_{x}(0,t)$, $U_{y}(0,t)$:

1) they are wide-sense stationary (WSS);

2) they are jointly Gaussian processes.\\

Regarding assumption 1), we plan to exploit the following extension
of result (\cite{papo}, p. 418, eq. (12-76)~):\\

\textbf{\emph{Theorem}}\textbf{ 1}

Consider the jointly WSS stochastic processes $x(t)$ and $y(t)$,
and let 
\[
\tilde{X}(f)\equiv\mathcal{F}[x(t)]=\int_{-\infty}^{\infty}x(t)\e^{-j2\pi ft}\mbox{d} t
\]
 the Fourier transform of $x$ ( in the mean-square (MS) sense), and
$\tilde{Y}(f)$ is similarly defined. Let their cross power spectral
density (PSD) be $G_{xy}(f)=\mathcal{F}[R_{xy}(\tau)]=\mathcal{F}[E[x(t+\tau)y^{*}(t)]]$.
Then
\begin{equation}
E[\tilde{X}(f)\tilde{Y}^{*}(u)]=G_{xy}(f)\delta(u-f)\qquad\square\label{eq:papoulis}
\end{equation}
\\

As a byproduct, we also have 
\[
E[\tilde{X}(f)\tilde{X}^{*}(u)]=G_{x}(f)\delta(u-f).
\]

This theorem thus shows that the Fourier transform of \emph{any MS-integrable
WSS process} is nonstationary white noise, and thus \emph{the spectral
lines of its Fourier transform are }\textbf{\emph{uncorrelated}}.
This was one of the key assumptions about the input field in \cite{Pontus},\cite{carena_JLT},
which therefore was tantamount to assuming a WSS input field in the
time domain.

Regarding assumption 2), we plan to exploit the following result,
known as the \emph{complex Gaussian moment theorem} (CGMT), a generalization
to complex variables of Isserlis theorem \cite{reed,goodman}:\\

\textbf{\emph{Theorem 2}}

Let $U_{1},U_{2},\ldots,U_{2k}$ be zero-mean jointly circular complex
Gaussian random variables. Then
\begin{equation}
E[U_{1}^{*}U_{2}^{*}...U_{k}^{*}U_{k+1}U_{k+2}...U_{2k}]=\sum_{\pi}E[U_{1}^{*}U_{p}]E[U_{2}^{*}U_{q}].....E[U_{k}^{*}U_{r}]\label{eq:isserlis}
\end{equation}
where $\sum_{\pi}$ denotes a summation over the $k!$ possible permutations
$(p,q,...,r)$ of indices $(k+1,k+2,...,2k)\qquad\square$\\

For instance, 
\begin{align}
E[U_{1}^{*}U_{2}^{*}U_{3}^{*}U_{4}U_{5}U_{6}] & =E[U_{1}^{*}U_{4}]E[U_{2}^{*}U_{5}]E[U_{3}^{*}U_{6}]\nonumber \\
 & +E[U_{1}^{*}U_{4}]E[U_{2}^{*}U_{6}]E[U_{3}^{*}U_{5}]\nonumber \\
 & +E[U_{1}^{*}U_{5}]E[U_{2}^{*}U_{4}]E[U_{3}^{*}U_{6}]\nonumber \\
 & +E[U_{1}^{*}U_{5}]E[U_{2}^{*}U_{6}]E[U_{3}^{*}U_{4}]\nonumber \\
 & +E[U_{1}^{*}U_{6}]E[U_{2}^{*}U_{4}]E[U_{3}^{*}U_{5}]\nonumber \\
 & +E[U_{1}^{*}U_{6}]E[U_{2}^{*}U_{5}]E[U_{3}^{*}U_{4}].\label{eq:iss_6}
\end{align}
(scheme: 1,2,3 containing the conjugate terms stay at their place.
Terms 4,5,6 get al possible permuted positions).\\

Let's now start the new proof. We are interested in the PSD $\hat{G}_{x,p}(f)$
of the NLI field $U_{x,p}(L,t)=\mathcal{F}^{-1}[\tilde{U}_{x,p}(L,f)]$.
By theorem 1 we have: 
\begin{equation}
E[\tilde{U}_{x,p}(L,f)\tilde{U}_{x,p}^{*}(L,u)]=\hat{G}_{x,p}(f)\delta(u-f).\label{eq:uno}
\end{equation}
and we wish to get an expression for $\hat{G}_{x,p}(f)$ from the
left hand side, which can be explicitly calculated using (\ref{eq:field_x}):
\begin{align}
 & \frac{E[\tilde{U}_{x,p}(L,f)\tilde{U}_{x,p}^{*}(L,u)]}{\Phi_{NL}^{2}}=\nonumber \\
 & E[\iint_{-\infty}^{\infty}\tilde{\eta}(f_{1}f_{2})[\tilde{U}_{x}(0,f+f_{1})\tilde{U}_{x}^{*}(0,f+f_{1}+f_{2})\tilde{U}_{x}(0,f+f_{2})+\nonumber \\
 & \tilde{U}_{x}(0,f+f_{1})\tilde{U}_{y}^{*}(0,f+f_{1}+f_{2})\tilde{U}_{y}(0,f+f_{2})]\mbox{d} f_{1}\mbox{d} f_{2}\cdot\nonumber \\
 & \iint_{-\infty}^{\infty}\tilde{\eta}(f_{3}f_{4})^{*}[\tilde{U}_{x}^{*}(0,u+f_{3})\tilde{U}_{x}(0,u+f_{3}+f_{4})\tilde{U}_{x}^{*}(0,u+f_{4})+\nonumber \\
 & \tilde{U}_{x}^{*}(0,u+f_{3})\tilde{U}_{y}(0,u+f_{3}+f_{4})\tilde{U}_{y}^{*}(0,u+f_{4})]\mbox{d} f_{3}\mbox{d} f_{4}]=\nonumber \\
 & \iiiint_{-\infty}^{\infty}\mbox{d} f_{1}\mbox{d} f_{2}\mbox{d} f_{3}\mbox{d} f_{4}\tilde{\eta}(f_{1}f_{2})\tilde{\eta}(f_{3}f_{4})^{*}\cdot\nonumber \\
 & \{E[\tilde{U}_{x}(0,f+f_{1})\tilde{U}_{x}^{*}(0,f+f_{1}+f_{2})\tilde{U}_{x}(0,f+f_{2})\tilde{U}_{x}^{*}(0,u+f_{3})\tilde{U}_{x}(0,u+f_{3}+f_{4})\tilde{U}_{x}^{*}(0,u+f_{4})]+\nonumber \\
 & E[\tilde{U}_{x}(0,f+f_{1})\tilde{U}_{y}^{*}(0,f+f_{1}+f_{2})\tilde{U}_{y}(0,f+f_{2})\tilde{U}_{x}^{*}(0,u+f_{3})\tilde{U}_{y}(0,u+f_{3}+f_{4})\tilde{U}_{y}^{*}(0,u+f_{4})]+\nonumber \\
 & E[\tilde{U}_{x}(0,f+f_{1})\tilde{U}_{x}^{*}(0,f+f_{1}+f_{2})\tilde{U}_{x}(0,f+f_{2})\tilde{U}_{x}^{*}(0,u+f_{3})\tilde{U}_{y}(0,u+f_{3}+f_{4})\tilde{U}_{y}^{*}(0,u+f_{4})]+\\
 & E[\tilde{U}_{x}(0,f+f_{1})\tilde{U}_{y}^{*}(0,f+f_{1}+f_{2})\tilde{U}_{y}(0,f+f_{2})\tilde{U}_{x}^{*}(0,u+f_{3})\tilde{U}_{x}(0,u+f_{3}+f_{4})\tilde{U}_{x}^{*}(0,u+f_{4})]\}.\label{eq:campo_1}
\end{align}

The above expectations of circular Gaussian RVs can now be obtained
by the following Theorem 3, which exploits Theorems 1 and 2 and is
proved in Appendix 1:

\textbf{Theorem 3} 

For jointly stationary circular complex Gaussian zero-mean processes
$A(t),B(t),C(t),D(t),E(t),F(t)$ we have the general formula
\begin{align}
E\left[\tilde{A}(f+f_{1})\tilde{B}^{*}(f+f_{1}+f_{2})\tilde{C}(f+f_{2})\tilde{D}^{*}(u+f_{3})\tilde{E}(u+f_{3}+f_{4})\tilde{F}^{*}(u+f_{4})\right] & =\nonumber \\
\Bigl[G_{ab}(f+f_{1})G_{cd}(f)G_{ef}(f+f_{4})\delta(f_{2})\delta(f_{3})+\nonumber \\
G_{ab}(f+f_{1})G_{ed}(f+f_{3})G_{cf}(f)\delta(f_{2})\delta(f_{4})+\nonumber \\
G_{cb}(f+f_{2})G_{ad}(f)G_{ef}(f+f_{4})\delta(f_{1})\delta(f_{3})+\nonumber \\
G_{cb}(f+f_{2})G_{ed}(f+f_{3})G_{af}(f)\delta(f_{1})\delta(f_{4})+\nonumber \\
G_{eb}(f+f_{1}+f_{2})G_{ad}(f+f_{1})G_{cf}(f+f_{2})\delta(f_{3}-f_{1})\delta(f_{4}-f_{2})+\nonumber \\
G_{eb}(f+f_{1}+f_{2})G_{cd}(f+f_{2})G_{af}(f+f_{1})\delta(f_{4}-f_{1})\delta(f_{3}-f_{2})\Bigr]\nonumber \\
\cdot\delta(u-f) & \qquad\square\label{eq:form_compact-1}
\end{align}

(scheme: b,d,f containing the conjugate terms stay at their place.
Terms a,c,e get al possible permuted positions. Arguments of terms
in products must be the same, hence the deltas) 

We next apply the general formula (\ref{eq:form_compact-1}) to the
expectations in (\ref{eq:campo_1}) to get:

First expectation:
\begin{align}
 & E\left[\underbrace{\tilde{U}_{x}(0,f+f_{1})}_{A}\underbrace{\tilde{U}_{x}^{*}(0,f+f_{1}+f_{2})}_{\mathbf{B}}\underbrace{\tilde{U}_{x}(0,f+f_{2})}_{C}\underbrace{\tilde{U}_{x}^{*}(0,u+f_{3})}_{\mathbf{D}}\underbrace{\tilde{U}_{x}(0,u+f_{3}+f_{4})}_{E}\underbrace{\tilde{U}_{x}^{*}(0,u+f_{4})}_{\mathbf{F}}\right]=\nonumber \\
 & \delta(u-f)\Bigl[\underbrace{G_{xx}(f+f_{1})}_{G_{ab}(f+f_{1})}\underbrace{G_{xx}(f)}_{G_{cd}(f)}\underbrace{G_{xx}(f+f_{4})}_{G_{ef}(f+f_{4})}\delta(f_{2})\delta(f_{3})+\nonumber \\
 & \underbrace{G_{xx}(f+f_{1})}_{G_{ab}(f+f_{1})}\underbrace{G_{xx}(f+f_{3})}_{G_{ed}(f+f_{3})}\underbrace{G_{xx}(f)}_{G_{cf}(f)}\delta(f_{2})\delta(f_{4})+\nonumber \\
 & \underbrace{G_{xx}(f+f_{2})}_{G_{cb}(f+f_{2})}\underbrace{G_{xx}(f)}_{G_{ad}(f)}\underbrace{G_{xx}(f+f_{4})}_{G_{ef}(f+f_{4})}\delta(f_{1})\delta(f_{3})+\nonumber \\
 & \underbrace{G_{xx}(f+f_{2})}_{G_{cb}(f+f_{2})}\underbrace{G_{xx}(f+f_{3})}_{G_{ed}(f+f_{3})}\underbrace{G_{xx}(f)}_{G_{af}(f)}\delta(f_{1})\delta(f_{4})+\nonumber \\
 & \underbrace{G_{xx}(f+f_{1}+f_{2})}_{G_{eb}(f+f_{1}+f_{2})}\underbrace{G_{xx}(f+f_{1})}_{G_{ad}(f+f_{1})}\underbrace{G_{xx}(f+f_{2})}_{G_{cf}(f+f_{2})}\delta(f_{3}-f_{1})\delta(f_{4}-f_{2})+\nonumber \\
 & \underbrace{G_{xx}(f+f_{1}+f_{2})}_{G_{eb}(f+f_{1}+f_{2})}\underbrace{G_{xx}(f+f_{2})}_{G_{cd}(f+f_{2})}\underbrace{G_{xx}(f+f_{1})}_{G_{af}(f+f_{1})}\delta(f_{4}-f_{1})\delta(f_{3}-f_{2})\Bigr]\label{eq:Eform_1}
\end{align}
where $G_{xx}\equiv\hat{G}_{x}$. 

Second expectation:
\begin{align*}
 & E\left[\underbrace{\tilde{U}_{x}(0,f+f_{1})}_{A}\underbrace{\tilde{U}_{y}^{*}(0,f+f_{1}+f_{2})}_{\mathbf{B}}\underbrace{\tilde{U}_{y}(0,f+f_{2})}_{C}\underbrace{\tilde{U}_{x}^{*}(0,u+f_{3})}_{\mathbf{D}}\underbrace{\tilde{U}_{y}(0,u+f_{3}+f_{4})}_{E}\underbrace{\tilde{U}_{y}^{*}(0,u+f_{4})}_{\mathbf{F}}\right]=\\
 & \delta(u-f)\Bigl[\underbrace{G_{xy}(f+f_{1})}_{G_{ab}(f+f_{1})}\underbrace{G_{yx}(f)}_{G_{cd}(f)}\underbrace{G_{yy}(f+f_{4})}_{G_{ef}(f+f_{4})}\delta(f_{2})\delta(f_{3})+\\
 & \underbrace{G_{xy}(f+f_{1})}_{G_{ab}(f+f_{1})}\underbrace{G_{yx}(f+f_{3})}_{G_{ed}(f+f_{3})}\underbrace{G_{yy}(f)}_{G_{cf}(f)}\delta(f_{2})\delta(f_{4})+\\
 & \underbrace{G_{yy}(f+f_{2})}_{G_{cb}(f+f_{2})}\underbrace{G_{xx}(f)}_{G_{ad}(f)}\underbrace{G_{yy}(f+f_{4})}_{G_{ef}(f+f_{4})}\delta(f_{1})\delta(f_{3})+\\
 & \underbrace{G_{yy}(f+f_{2})}_{G_{cb}(f+f_{2})}\underbrace{G_{yx}(f+f_{3})}_{G_{ed}(f+f_{3})}\underbrace{G_{xy}(f)}_{G_{af}(f)}\delta(f_{1})\delta(f_{4})+\\
 & \underbrace{G_{yy}(f+f_{1}+f_{2})}_{G_{eb}(f+f_{1}+f_{2})}\underbrace{G_{xx}(f+f_{1})}_{G_{ad}(f+f_{1})}\underbrace{G_{yy}(f+f_{2})}_{G_{cf}(f+f_{2})}\delta(f_{3}-f_{1})\delta(f_{4}-f_{2})+\\
 & \underbrace{G_{yy}(f+f_{1}+f_{2})}_{G_{eb}(f+f_{1}+f_{2})}\underbrace{G_{yx}(f+f_{2})}_{G_{cd}(f+f_{2})}\underbrace{G_{xy}(f+f_{1})}_{G_{af}(f+f_{1})}\delta(f_{4}-f_{1})\delta(f_{3}-f_{2})\Bigr]
\end{align*}
where $G_{yy}\equiv\hat{G}_{y}$, and assuming uncorrelated X and
Y (i.e., $G_{yx}\equiv0$) we get
\begin{align}
 & E\left[\tilde{U}_{x}(0,f+f_{1})\tilde{U}_{y}^{*}(0,f+f_{1}+f_{2})\tilde{U}_{y}(0,f+f_{2})\tilde{U}_{x}^{*}(0,u+f_{3})\tilde{U}_{y}(0,u+f_{3}+f_{4})\tilde{U}_{y}^{*}(0,u+f_{4})\right]=\nonumber \\
 & \delta(u-f)\Bigl[G_{yy}(f+f_{2})G_{xx}(f)G_{yy}(f+f_{4})\delta(f_{1})\delta(f_{3})+\nonumber \\
 & G_{yy}(f+f_{1}+f_{2})G_{xx}(f+f_{1})G_{yy}(f+f_{2})\delta(f_{3}-f_{1})\delta(f_{4}-f_{2})\Bigr].\label{eq:Eform_2}
\end{align}

Third expectation:
\begin{align*}
 & E\left[\underbrace{\tilde{U}_{x}(0,f+f_{1})}_{A}\underbrace{\tilde{U}_{x}^{*}(0,f+f_{1}+f_{2})}_{\mathbf{B}}\underbrace{\tilde{U}_{x}(0,f+f_{2})}_{C}\underbrace{\tilde{U}_{x}^{*}(0,u+f_{3})}_{\mathbf{D}}\underbrace{\tilde{U}_{y}(0,u+f_{3}+f_{4})}_{E}\underbrace{\tilde{U}_{y}^{*}(0,u+f_{4})}_{\mathbf{F}}\right]=\\
 & \delta(u-f)\Bigl[\underbrace{G_{xx}(f+f_{1})}_{G_{ab}(f+f_{1})}\underbrace{G_{xx}(f)}_{G_{cd}(f)}\underbrace{G_{yy}(f+f_{4})}_{G_{ef}(f+f_{4})}\delta(f_{2})\delta(f_{3})+\\
 & \underbrace{G_{xx}(f+f_{1})}_{G_{ab}(f+f_{1})}\underbrace{G_{yx}(f+f_{3})}_{G_{ed}(f+f_{3})}\underbrace{G_{xy}(f)}_{G_{cf}(f)}\delta(f_{2})\delta(f_{4})+\\
 & \underbrace{G_{xx}(f+f_{2})}_{G_{cb}(f+f_{2})}\underbrace{G_{xx}(f)}_{G_{ad}(f)}\underbrace{G_{yy}(f+f_{4})}_{G_{ef}(f+f_{4})}\delta(f_{1})\delta(f_{3})+\\
 & \underbrace{G_{xx}(f+f_{2})}_{G_{cb}(f+f_{2})}\underbrace{G_{yx}(f+f_{3})}_{G_{ed}(f+f_{3})}\underbrace{G_{xy}(f)}_{G_{af}(f)}\delta(f_{1})\delta(f_{4})+\\
 & \underbrace{G_{yx}(f+f_{1}+f_{2})}_{G_{eb}(f+f_{1}+f_{2})}\underbrace{G_{xx}(f+f_{1})}_{G_{ad}(f+f_{1})}\underbrace{G_{xy}(f+f_{2})}_{G_{cf}(f+f_{2})}\delta(f_{3}-f_{1})\delta(f_{4}-f_{2})+\\
 & \underbrace{G_{yx}(f+f_{1}+f_{2})}_{G_{eb}(f+f_{1}+f_{2})}\underbrace{G_{xx}(f+f_{2})}_{G_{cd}(f+f_{2})}\underbrace{G_{xy}(f+f_{1})}_{G_{af}(f+f_{1})}\delta(f_{4}-f_{1})\delta(f_{3}-f_{2})\Bigr]
\end{align*}

and assuming uncorrelated X and Y (i.e., $G_{yx}\equiv0$) we get
\begin{align}
 & E\left[\tilde{U}_{x}(0,f+f_{1})\tilde{U}_{x}^{*}(0,f+f_{1}+f_{2})\tilde{U}_{x}(0,f+f_{2})\tilde{U}_{x}^{*}(0,u+f_{3})\tilde{U}_{y}(0,u+f_{3}+f_{4})\tilde{U}_{y}^{*}(0,u+f_{4})\right]=\nonumber \\
 & \delta(u-f)\Bigl[G_{xx}(f+f_{1})G_{xx}(f)G_{yy}(f+f_{4})\delta(f_{2})\delta(f_{3})+\nonumber \\
 & G_{xx}(f+f_{2})G_{xx}(f)G_{yy}(f+f_{4})\delta(f_{1})\delta(f_{3})\Bigr].\label{eq:Eform_3}
\end{align}

Fourth expectation:
\begin{align*}
 & E\left[\underbrace{\tilde{U}_{x}(0,f+f_{1})}_{A}\underbrace{\tilde{U}_{y}^{*}(0,f+f_{1}+f_{2})}_{\mathbf{B}}\underbrace{\tilde{U}_{y}(0,f+f_{2})}_{C}\underbrace{\tilde{U}_{x}^{*}(0,u+f_{3})}_{\mathbf{D}}\underbrace{\tilde{U}_{x}(0,u+f_{3}+f_{4})}_{E}\underbrace{\tilde{U}_{x}^{*}(0,u+f_{4})}_{\mathbf{F}}\right]=\\
 & \delta(u-f)\Bigl[\underbrace{G_{xy}(f+f_{1})}_{G_{ab}(f+f_{1})}\underbrace{G_{yx}(f)}_{G_{cd}(f)}\underbrace{G_{xx}(f+f_{4})}_{G_{ef}(f+f_{4})}\delta(f_{2})\delta(f_{3})+\\
 & \underbrace{G_{xy}(f+f_{1})}_{G_{ab}(f+f_{1})}\underbrace{G_{xx}(f+f_{3})}_{G_{ed}(f+f_{3})}\underbrace{G_{yx}(f)}_{G_{cf}(f)}\delta(f_{2})\delta(f_{4})+\\
 & \underbrace{G_{yy}(f+f_{2})}_{G_{cb}(f+f_{2})}\underbrace{G_{xx}(f)}_{G_{ad}(f)}\underbrace{G_{xx}(f+f_{4})}_{G_{ef}(f+f_{4})}\delta(f_{1})\delta(f_{3})+\\
 & \underbrace{G_{yy}(f+f_{2})}_{G_{cb}(f+f_{2})}\underbrace{G_{xx}(f+f_{3})}_{G_{ed}(f+f_{3})}\underbrace{G_{xx}(f)}_{G_{af}(f)}\delta(f_{1})\delta(f_{4})+\\
 & \underbrace{G_{xy}(f+f_{1}+f_{2})}_{G_{eb}(f+f_{1}+f_{2})}\underbrace{G_{xx}(f+f_{1})}_{G_{ad}(f+f_{1})}\underbrace{G_{yx}(f+f_{2})}_{G_{cf}(f+f_{2})}\delta(f_{3}-f_{1})\delta(f_{4}-f_{2})+\\
 & \underbrace{G_{xy}(f+f_{1}+f_{2})}_{G_{eb}(f+f_{1}+f_{2})}\underbrace{G_{yx}(f+f_{2})}_{G_{cd}(f+f_{2})}\underbrace{G_{xx}(f+f_{1})}_{G_{af}(f+f_{1})}\delta(f_{4}-f_{1})\delta(f_{3}-f_{2})\Bigr]
\end{align*}

and assuming uncorrelated X and Y (i.e., $G_{yx}\equiv0$) we get
\begin{align}
 & E\left[\tilde{U}_{x}(0,f+f_{1})\tilde{U}_{y}^{*}(0,f+f_{1}+f_{2})\tilde{U}_{y}(0,f+f_{2})\tilde{U}_{x}^{*}(0,u+f_{3})\tilde{U}_{x}(0,u+f_{3}+f_{4})\tilde{U}_{x}^{*}(0,u+f_{4})\right]=\nonumber \\
 & \delta(u-f)\Bigl[G_{yy}(f+f_{2})G_{xx}(f)G_{xx}(f+f_{4})\delta(f_{1})\delta(f_{3})+\nonumber \\
 & G_{yy}(f+f_{2})G_{xx}(f+f_{3})G_{xx}(f)\delta(f_{1})\delta(f_{4})\Bigr].\label{eq:Eform_3-1}
\end{align}

Substitution of (\ref{eq:Eform_1}),(\ref{eq:Eform_2}),(\ref{eq:Eform_3}),(\ref{eq:Eform_3-1})
into (\ref{eq:campo_1}) finally gives 
\begin{align*}
 & \frac{E[\tilde{U}_{x,p}(L,f)\tilde{U}_{x,p}^{*}(L,u)]}{\Phi_{NL}^{2}}=\delta(u-f)\{\iiiint_{-\infty}^{\infty}\mbox{d} f_{1}\mbox{d} f_{2}\mbox{d} f_{3}\mbox{d} f_{4}\tilde{\eta}(f_{1}f_{2})\tilde{\eta}(f_{3}f_{4})^{*}\cdot\\
 & \cdot\{\Bigl[\hat{G}_{x}(f+f_{1})\hat{G}_{x}(f)\hat{G}_{x}(f+f_{4})\delta(f_{2})\delta(f_{3})+\hat{G}_{x}(f+f_{1})\hat{G}_{x}(f+f_{3})\hat{G}_{x}(f)\delta(f_{2})\delta(f_{4})+\\
 & \hat{G}_{x}(f+f_{2})\hat{G}_{x}(f)\hat{G}_{x}(f+f_{4})\delta(f_{1})\delta(f_{3})+\hat{G}_{x}(f+f_{2})\hat{G}_{x}(f+f_{3})\hat{G}_{x}(f)\delta(f_{1})\delta(f_{4})+\\
 & \!\!\!\!\!\!\hat{\!\!\!G}_{x}(f+f_{1}+f_{2})\!\hat{G}_{x}(f+f_{1})\negthinspace\hat{G}_{x}(f+f_{2})\delta(f_{3}-f_{1})\delta(f_{4}-f_{2})\!+\!\hat{G}_{x}(f+f_{1}+f_{2})\!\hat{G}_{x}(f+f_{2})\!\hat{G}_{x}(f+f_{1})\delta(f_{4}-f_{1})\delta(f_{3}-f_{2})\Bigr]\!+\\
 & \Bigl[\hat{G}_{y}(f+f_{2})\hat{G}_{x}(f)\hat{G}_{y}(f+f_{4})\delta(f_{1})\delta(f_{3})+\hat{G}_{y}(f+f_{1}+f_{2})\hat{G}_{x}(f+f_{1})\hat{G}_{y}(f+f_{2})\delta(f_{3}-f_{1})\delta(f_{4}-f_{2})\Bigr]+\\
 & \Bigl[\hat{G}_{x}(f+f_{1})\hat{G}_{x}(f)\hat{G}_{y}(f+f_{4})\delta(f_{2})\delta(f_{3})+\hat{G}_{x}(f+f_{2})\hat{G}_{x}(f)\hat{G}_{y}(f+f_{4})\delta(f_{1})\delta(f_{3})\Bigr]+\\
 & \Bigl[\hat{G}_{y}(f+f_{2})\hat{G}_{x}(f)\hat{G}_{x}(f+f_{4})\delta(f_{1})\delta(f_{3})+\hat{G}_{y}(f+f_{2})\hat{G}_{x}(f+f_{3})\hat{G}_{x}(f)\delta(f_{1})\delta(f_{4})\Bigr]\}\}.
\end{align*}

From (\ref{eq:uno}), the term multiplying $\delta(u-f)$ must be
the desired PSD. 

Each of the 6 lines in the above quadruple integral provides 2 integrands
whose integral we must calculate:\\
1.1)
\begin{eqnarray*}
\iiiint_{-\infty}^{\infty}\mbox{d} f_{1}\mbox{d} f_{2}\mbox{d} f_{3}\mbox{d} f_{4}\tilde{\eta}(f_{1}f_{2})\tilde{\eta}(f_{3}f_{4})^{*}\hat{G}_{x}(f+f_{1})\hat{G}_{x}(f)\hat{G}_{x}(f+f_{4})\delta(f_{2})\delta(f_{3}) & =\\
\hat{G}_{x}(f)\left[\iint\tilde{\eta}(f_{1}f_{2})\hat{G}_{x}(f+f_{1})\delta(f_{2})\mbox{d} f_{1}\mbox{d} f_{2}\right]\left[\iint\tilde{\eta}(f_{3}f_{4})^{*}\hat{G}_{x}(f+f_{4})\delta(f_{3})\mbox{d} f_{3}\mbox{d} f_{4}\right] & =\\
\hat{G}_{x}(f)\left[\int\hat{G}_{x}(f+f_{1})\underbrace{\left(\int\tilde{\eta}(f_{1}f_{2})\delta(f_{2})\mbox{d} f_{2}\right)}_{\tilde{\eta}(0)\cdot\int\delta(f_{2})\mbox{d} f_{2}}\mbox{d} f_{1}\right]\left[\int\hat{G}_{x}(f+f_{4})\underbrace{\left(\int\tilde{\eta}(f_{3}f_{4})^{*}\delta(f_{3})\mbox{d} f_{3}\right)}_{\tilde{\eta}^{*}(0)\cdot\int\delta(f_{3})\mbox{d} f_{3}}\mbox{d} f_{4}\right] & =\\
\hat{G}_{x}(f)|\tilde{\eta}(0)|^{2}\left[\int\hat{G}_{x}(f+f_{1})\mbox{d} f_{1}\right]\left[\int\hat{G}_{x}(f+f_{4})\mbox{d} f_{4}\right] & =\\
\hat{G}_{x}(f)|\tilde{\eta}(0)|^{2}\frac{P_{x}}{P_{0}}
\end{eqnarray*}
\\
1.2)
\begin{eqnarray*}
\iiiint_{-\infty}^{\infty}\mbox{d} f_{1}\mbox{d} f_{2}\mbox{d} f_{3}\mbox{d} f_{4}\tilde{\eta}(f_{1}f_{2})\tilde{\eta}(f_{3}f_{4})^{*}\hat{G}_{x}(f+f_{1})\hat{G}_{x}(f+f_{3})\hat{G}_{x}(f)\delta(f_{2})\delta(f_{4}) & =\\
\hat{G}_{x}(f)\left[\iint\tilde{\eta}(f_{1}f_{2})\hat{G}_{x}(f+f_{1})\delta(f_{2})\mbox{d} f_{1}\mbox{d} f_{2}\right]\left[\iint\tilde{\eta}(f_{3}f_{4})^{*}\hat{G}_{x}(f+f_{3})\delta(f_{4})\mbox{d} f_{3}\mbox{d} f_{4}\right] & =\\
\hat{G}_{x}(f)\left[\int\hat{G}_{x}(f+f_{1})\underbrace{\left(\int\tilde{\eta}(f_{1}f_{2})\delta(f_{2})\mbox{d} f_{2}\right)}_{\tilde{\eta}(0)\cdot\int\delta(f_{2})\mbox{d} f_{2}}\mbox{d} f_{1}\right]\left[\int\hat{G}_{x}(f+f_{3})\underbrace{\left(\int\tilde{\eta}(f_{3}f_{4})^{*}\delta(f_{4})\mbox{d} f_{4}\right)}_{\tilde{\eta}^{*}(0)\cdot\int\delta(f_{4})\mbox{d} f_{4}}\mbox{d} f_{3}\right] & =\\
\hat{G}_{x}(f)|\tilde{\eta}(0)|^{2}\left[\int\hat{G}_{x}(f+f_{1})\mbox{d} f_{1}\right]\left[\int\hat{G}_{x}(f+f_{3})\mbox{d} f_{3}\right] & =\\
\hat{G}_{x}(f)|\tilde{\eta}(0)|^{2}\frac{P_{x}}{P_{0}}
\end{eqnarray*}
\\
2.1)
\begin{eqnarray*}
\iiiint_{-\infty}^{\infty}\mbox{d} f_{1}\mbox{d} f_{2}\mbox{d} f_{3}\mbox{d} f_{4}\tilde{\eta}(f_{1}f_{2})\tilde{\eta}(f_{3}f_{4})^{*}\hat{G}_{x}(f+f_{2})\hat{G}_{x}(f)\hat{G}_{x}(f+f_{4})\delta(f_{1})\delta(f_{3}) & =\\
\hat{G}_{x}(f)\left[\iint\tilde{\eta}(f_{1}f_{2})\hat{G}_{x}(f+f_{2})\delta(f_{1})\mbox{d} f_{1}\mbox{d} f_{2}\right]\left[\iint\tilde{\eta}(f_{3}f_{4})^{*}\hat{G}_{x}(f+f_{4})\delta(f_{3})\mbox{d} f_{3}\mbox{d} f_{4}\right] & =\\
\hat{G}_{x}(f)\left[\int\hat{G}_{x}(f+f_{2})\underbrace{\left(\int\tilde{\eta}(f_{1}f_{2})\delta(f_{1})\mbox{d} f_{1}\right)}_{\tilde{\eta}(0)}\mbox{d} f_{2}\right]\left[\int\hat{G}_{x}(f+f_{4})\underbrace{\left(\int\tilde{\eta}(f_{3}f_{4})^{*}\delta(f_{3})\mbox{d} f_{3}\right)}_{\tilde{\eta}^{*}(0)}\mbox{d} f_{4}\right] & =\\
\hat{G}_{x}(f)|\tilde{\eta}(0)|^{2}\left[\int\hat{G}_{x}(f+f_{2})\mbox{d} f_{2}\right]\left[\int\hat{G}_{x}(f+f_{4})\mbox{d} f_{4}\right] & =\\
\hat{G}_{x}(f)|\tilde{\eta}(0)|^{2}\frac{P_{x}}{P_{0}}
\end{eqnarray*}
\\
2.2) 
\begin{eqnarray*}
\iiiint_{-\infty}^{\infty}\mbox{d} f_{1}\mbox{d} f_{2}\mbox{d} f_{3}\mbox{d} f_{4}\tilde{\eta}(f_{1}f_{2})\tilde{\eta}(f_{3}f_{4})^{*}\hat{G}_{x}(f+f_{2})\hat{G}_{x}(f+f_{3})\hat{G}_{x}(f)\delta(f_{1})\delta(f_{4}) & =\\
\hat{G}_{x}(f)\left[\iint\tilde{\eta}(f_{1}f_{2})\hat{G}_{x}(f+f_{2})\delta(f_{1})\mbox{d} f_{1}\mbox{d} f_{2}\right]\left[\iint\tilde{\eta}(f_{3}f_{4})^{*}\hat{G}_{x}(f+f_{3})\delta(f_{4})\mbox{d} f_{3}\mbox{d} f_{4}\right] & =\\
\hat{G}_{x}(f)\left[\int\hat{G}_{x}(f+f_{2})\underbrace{\left(\int\tilde{\eta}(f_{1}f_{2})\delta(f_{1})\mbox{d} f_{1}\right)}_{\tilde{\eta}(0)}\mbox{d} f_{2}\right]\left[\int\hat{G}_{x}(f+f_{3})\underbrace{\left(\int\tilde{\eta}(f_{3}f_{4})^{*}\delta(f_{4})\mbox{d} f_{4}\right)}_{\tilde{\eta}^{*}(0)}\mbox{d} f_{3}\right] & =\\
\hat{G}_{x}(f)|\tilde{\eta}(0)|^{2}\left[\int\hat{G}_{x}(f+f_{2})\mbox{d} f_{2}\right]\left[\int\hat{G}_{x}(f+f_{3})\mbox{d} f_{3}\right] & =\\
\hat{G}_{x}(f)|\tilde{\eta}(0)|^{2}\frac{P_{x}}{P_{0}}
\end{eqnarray*}
\\
3.1)
\begin{eqnarray*}
\iiiint_{-\infty}^{\infty}\mbox{d} f_{1}\mbox{d} f_{2}\mbox{d} f_{3}\mbox{d} f_{4}\tilde{\eta}(f_{1}f_{2})\tilde{\eta}(f_{3}f_{4})^{*}\hat{G}_{x}(f+f_{1}+f_{2})\hat{G}_{x}(f+f_{1})\hat{G}_{x}(f+f_{2})\delta(f_{3}-f_{1})\delta(f_{4}-f_{2}) & =\\
\iint\tilde{\eta}(f_{1}f_{2})\hat{G}_{x}(f+f_{1}+f_{2})\hat{G}_{x}(f+f_{1})\hat{G}_{x}(f+f_{2})\left[\iint\tilde{\eta}(f_{3}f_{4})^{*}\delta(f_{3}-f_{1})\delta(f_{4}-f_{2})\mbox{d} f_{3}\mbox{d} f_{4}\right]\mbox{d} f_{1}\mbox{d} f_{2} & =\\
\iint\tilde{\eta}(f_{1}f_{2})\hat{G}_{x}(f+f_{1}+f_{2})\hat{G}_{x}(f+f_{1})\hat{G}_{x}(f+f_{2})\underbrace{\left[\int\delta(f_{3}-f_{1})\underbrace{\int\tilde{\eta}(f_{3}f_{4})^{*}\delta(f_{4}-f_{2})\mbox{d} f_{4}}_{\tilde{\eta}^{*}(f_{3}f_{2})}\mbox{d} f_{3}\right]}_{\tilde{\eta}^{*}(f_{1}f_{2})}\mbox{d} f_{1}\mbox{d} f_{2} & =\\
\iint|\tilde{\eta}(f_{1}f_{2})|^{2}\hat{G}_{x}(f+f_{1}+f_{2})\hat{G}_{x}(f+f_{1})\hat{G}_{x}(f+f_{2})\mbox{d} f_{1}\mbox{d} f_{2}
\end{eqnarray*}

Each pair of delta removes two integrals, so that the PSD turns out
to be (first two lines above produce first 4 lines, third line above
produces 5th line, 4th line above produces 6th and 7th lines, and
final line above produces the last two lines):
\begin{align*}
\frac{\hat{G}_{x,p}(f)}{\Phi_{NL}^{2}} & =|\tilde{\eta}(0)|^{2}{\color{green}\iintop_{-\infty}^{\infty}\hat{G}_{x}(f+f_{1})\hat{G}_{x}(f)\hat{G}_{x}(f+f_{4})\mbox{d} f_{1}\mbox{d} f_{4}}\\
 & +|\tilde{\eta}(0)|^{2}{\color{green}\iintop_{-\infty}^{\infty}\hat{G}_{x}(f+f_{1})\hat{G}_{x}(f+f_{3})\hat{G}_{x}(f)\mbox{d} f_{1}\mbox{d} f_{3}}\\
 & +|\tilde{\eta}(0)|^{2}{\color{green}\iintop_{-\infty}^{\infty}\hat{G}_{x}(f+f_{2})\hat{G}_{x}(f)\hat{G}_{x}(f+f_{4})\mbox{d} f_{2}\mbox{d} f_{4}}\\
 & +|\tilde{\eta}(0)|^{2}{\color{green}\iintop_{-\infty}^{\infty}\hat{G}_{x}(f+f_{2})\hat{G}_{x}(f+f_{3})\hat{G}_{x}(f)\mbox{d} f_{2}\mbox{d} f_{3}}\\
 & +2\iintop_{-\infty}^{\infty}|\tilde{\eta}(f_{1}f_{2})|^{2}\hat{G}_{x}(f+f_{1}+f_{2})\hat{G}_{x}(f+f_{1})\hat{G}_{x}(f+f_{2})\mbox{d} f_{1}\mbox{d} f_{2}\\
 & +|\tilde{\eta}(0)|^{2}{\color{red}\iintop_{-\infty}^{\infty}\hat{G}_{y}(f+f_{2})\hat{G}_{x}(f)\hat{G}_{y}(f+f_{4})\mbox{d} f_{2}\mbox{d} f_{4}}\\
 & +\iintop_{-\infty}^{\infty}|\tilde{\eta}(f_{1}f_{2})|^{2}\hat{G}_{y}(f+f_{1}+f_{2})\hat{G}_{x}(f+f_{1})\hat{G}_{y}(f+f_{2})\mbox{d} f_{1}\mbox{d} f_{2}\\
 & +2\left({\color{magenta}|\tilde{\eta}(0)|^{2}\iintop_{-\infty}^{\infty}\hat{G}_{x}(f+f_{1})\hat{G}_{x}(f)\hat{G}_{y}(f+f_{4})\mbox{d} f_{1}\mbox{d} f_{4}}\right.+\\
 & \,\,\,\,\,\,\left.{\color{magenta}|\tilde{\eta}(0)|^{2}\iintop_{-\infty}^{\infty}\hat{G}_{x}(f+f_{2})\hat{G}_{x}(f)\hat{G}_{y}(f+f_{4})\mbox{d} f_{2}\mbox{d} f_{4}}\right).
\end{align*}

In summary, considering that by construction $\tilde{\eta}(0)=1$,
we have:
\begin{align}
\frac{\hat{G}_{x,p}(f)}{\Phi_{NL}^{2}} & =2\iintop_{-\infty}^{\infty}|\tilde{\eta}(f_{1}f_{2})|^{2}\hat{G}_{x}(f+f_{1}+f_{2})\hat{G}_{x}(f+f_{1})\hat{G}_{x}(f+f_{2})\mbox{d} f_{1}\mbox{d} f_{2}\nonumber \\
 & +\iintop_{-\infty}^{\infty}|\tilde{\eta}(f_{1}f_{2})|^{2}\hat{G}_{y}(f+f_{1}+f_{2})\hat{G}_{x}(f+f_{1})\hat{G}_{y}(f+f_{2})\mbox{d} f_{1}\mbox{d} f_{2}\nonumber \\
 & +\hat{G}_{x}(f)\left({\color{green}4}(\frac{P_{x}}{P_{0}})^{2}+{\color{magenta}4}\frac{P_{x}}{P_{0}}\frac{P_{y}}{P_{0}}+{\color{red}1}(\frac{P_{y}}{P_{0}})^{2}\right)\label{eq:finale1}
\end{align}

which confirms Johannisson's equation (\ref{eq:G_xp-2}) and completes
the desired alternative proof. 

\section{Intermediate Summary}

We have presented an alternative derivation of Johannisson's et al.
\cite{Pontus} result based on the new method in \cite{arxiv}. We
first remark that both the result in \cite{Pontus} and our new approach
are able to deal with correlated X and Y, although this feature was
not exploited here. Next we note that we did not have to assume independent
input spectral lines: this comes naturally from the stationarity of
the input process. Finally, the truly critical assumption in the model
in \cite{Pontus,carena_JLT} is therefore the assumption of Gaussianity
at any $z$ during propagation, which is implicit in the assumption
of a Gaussian input process, and the fact that the ``forcing terms''
in the RP equation are the linearly distorted signals at any $z$,
which thus remain Gaussian. 

Therefore the true limit of the GN model in \cite{Pontus,carena_JLT}
is that indeed starting from a non-Gaussian spectrum such as the one
of a digitally modulated signal\footnote{Although the authors in \cite{carena_JLT} present in their Appendix
B an appealing heuristic justification of their Gaussian signal assumption,
still their invoking the central limit theorem at their equation (37)
is not rigorous. They would conclude that any digitally modulated
signal with any number of levels has a Gaussian Fourier transform
(which in turn implies the time-domain signal itself is Gaussian),
which is clearly \emph{not} the case. }, it takes some finite propagation in a non-infinite dispersion line
to approximately get both a Gaussian spectrum and a Gaussian-like
time-domain signal. 

\section{Role of the last line in (\ref{eq:finale1})}

We now focus the attention of the last ``phase term'' in the main
result of \cite{Pontus}, i.e. our equation (\ref{eq:finale1}). We
note that $\int_{-\infty}^{\infty}\hat{G}_{x}(f)df=P_{x}/P_{0}$,
$\int_{-\infty}^{\infty}\hat{G}_{y}(f)df=P_{y}/P_{0}$, hence the
last term is
\begin{eqnarray*}
\left({\color{green}4}(\int_{-\infty}^{\infty}\hat{G}_{x}(f)df)^{2}+{\color{magenta}4}\int_{-\infty}^{\infty}\hat{G}_{x}(f)df\int_{-\infty}^{\infty}\hat{G}_{y}(f)df+{\color{red}1}(\int_{-\infty}^{\infty}\hat{G}_{y}(f)df)^{2}\right) & =\\
(2\frac{P_{x}}{P_{0}})^{2}+2(2\frac{P_{x}}{P_{0}})\frac{P_{y}}{P_{0}}+(\frac{P_{y}}{P_{0}})^{2} & =\\
(\frac{2P_{x}+P_{y}}{P_{0}})^{2}
\end{eqnarray*}
Going back to un-normalized kernels and PSDs, we can finally write
the sought nonlinear interference PSD as

\begin{align}
G_{x,p}(f) & =\mathcal{K}(0)^{2}\{2\iintop_{-\infty}^{\infty}|\tilde{\eta}(f_{1}f_{2})|^{2}G_{x}(f+f_{1}+f_{2})G_{x}(f+f_{1})G_{x}(f+f_{2})\mbox{d} f_{1}\mbox{d} f_{2}\nonumber \\
 & +\iintop_{-\infty}^{\infty}|\tilde{\eta}(f_{1}f_{2})|^{2}G_{y}(f+f_{1}+f_{2})G_{x}(f+f_{1})G_{y}(f+f_{2})\mbox{d} f_{1}\mbox{d} f_{2}\}\nonumber \\
 & +\mathcal{K}(0)^{2}\left(2P_{x}+P_{y}\right)^{2}G_{x}(f)\}.\label{eq:pontus_X_sol}
\end{align}

The last line can be interpreted as the PSD of the first-order regular
perturbation (RP1) solution of a system whose ``exact'' output is:
\begin{align*}
U_{x}(L,t)=U_{x}(0,t)\e^{-j\gamma\int_{0}^{L}\mathcal{G}(s)\mbox{d} s\cdot(2P_{x}+P_{y})}\cong & U_{x}(0,t)(1-j\overbrace{\gamma\int_{0}^{L}\mathcal{G}(s)\mbox{d} s}^{\mathcal{K}(0)}\cdot(2P_{x}+P_{y}))\\
= & U_{x}(0,t)-\underbrace{j\mathcal{K}(0)(2P_{x}+P_{y})U_{x}(0,t)}_{\triangleq U_{x,p}(L,t)}
\end{align*}
so that the PSD of \foreignlanguage{english}{$U_{x,p}(L,t)$ is $\mathcal{K}(0)^{2}(2P_{x}+P_{y})^{2}G_{x}(f)$.
Note the surprising average nonlinear phase rotation by an effective
power $2P_{x}+P_{y}$ which is larger than the actual average power
in the fiber $P_{x}+P_{y}$! This is one of the new unexpected features
of dispersion uncompensated systems. }

In such DU systems we are thus naturally lead to postulate an ansatz
of the DP dispersion managed nonlinear Schroedinger equation (DMNLSE)
of the kind 
\[
\tilde{\boldsymbol{A}}(z,f)=\sqrt{P_{0}\mathcal{G}(z)}\e^{j\frac{C(z)\omega^{2}}{2}}\left[\begin{array}{c}
\tilde{U}_{x}(z,f){\color{red}\e}^{{\color{red}-j\gamma\int_{0}^{L}\mathcal{G}(s)\mbox{d} s\cdot(2P_{x}+P_{y})}}\\
\tilde{U}_{y}(z,f){\color{red}\e}^{{\color{red}-j\gamma\int_{0}^{L}\mathcal{G}(s)\mbox{d} s\cdot(2P_{y}+P_{x})}}
\end{array}\right]
\]
which we call the dual-polarization \textbf{enhanced} regular perturbation
(DP-ERP) ansatz (from a similar ERP definition in \cite{RP}). We
prove below as an exercise that this change of variables removes the
unwanted spectral components in the last line of equation (\ref{eq:pontus_X_sol}).
That's what tacitly also Turin's group does in their detailed GN model
derivation \cite{poggio_GN_arxiv}.

\section{Analysis }

The procedure reported below re-derives from scratch the DM-NLSE and
the RP1 solution in dual polarization and in the correct phase-reference.\\

The\foreignlanguage{english}{ Manakov Nonlinear Schroedinger equation
(M-NLSE) }in engineering notation writes in the frequency domain as\foreignlanguage{english}{
\cite{OE2012}}

\begin{align}
\frac{\partial\tilde{\boldsymbol{A}}(z,f)}{\partial z}= & \frac{g(z)-j\omega^{2}\beta_{2}(z)}{2}\tilde{\boldsymbol{A}}(z,f)+\label{eq:Sch_eq_f-1}\\
 & -j\gamma(z)\iint_{-\infty}^{\infty}\underbrace{\tilde{\boldsymbol{A}}(z,f+f_{1})\tilde{\boldsymbol{A}}^{\dagger}(z,f+f_{1}+f_{2})\tilde{\boldsymbol{A}}(z,f+f_{2})}_{\left[\begin{array}{c}
(\tilde{A}_{x}^{*}(z,f+f_{1}+f_{2})\tilde{A}_{x}(z,f+f_{2})+\tilde{A}_{y}^{*}\tilde{A}_{y})\tilde{A}_{x}(z,f+f_{1})\\
(\tilde{A}_{x}^{*}(z,f+f_{1}+f_{2})\tilde{A}_{x}(z,f+f_{2})+\tilde{A}_{y}^{*}\tilde{A}_{y})\tilde{A}_{y}(z,f+f_{1})
\end{array}\right]}\mbox{d} f_{1}\mbox{d} f_{2}\nonumber 
\end{align}
where $\omega=2\pi f$, and $f$ is the frequency normalized to the
baud rate. The input modulated field $\tilde{\boldsymbol{A}}_{0}(f)$
may be pre-chirped to give $\tilde{\boldsymbol{A}}(0,f)\equiv\tilde{\boldsymbol{A}}_{0}(f)\mbox{e}^{j\frac{\omega^{2}}{2}\xi_{pre}}$,
where $\xi_{pre}$ is the pre-compensation. Now make the change of
variable 
\begin{equation}
\left[\begin{array}{c}
\tilde{A}_{x}\\
\tilde{A}_{y}
\end{array}\right](z,f)=\sqrt{P_{0}}\left[\begin{array}{c}
\tilde{U}_{x}(z,f)\e^{\frac{\ln\mathcal{G}(z)+jC(z)\omega^{2}}{2}-j\int_{0}^{z}\gamma(s)\mathcal{G}(s)\mbox{d} s\cdot(2P_{x}+P_{y})}\\
\tilde{U}_{y}(z,f)\e^{\frac{\ln\mathcal{G}(z)+jC(z)\omega^{2}}{2}-j\int_{0}^{z}\gamma(s)\mathcal{G}(s)\mbox{d} s\cdot(2P_{y}+P_{x})}
\end{array}\right]\label{change_1}
\end{equation}
where $P_{0}$ is a reference normalizing power, $\mathcal{G}(z)=\e^{\int_{0}^{z}g(s)\mbox{d} s}$
is the power gain at $z$, and $C(z)\triangleq\xi_{pre}-\int_{0}^{z}\beta_{2}(s)\mbox{d} s$
is the\emph{ cumulated dispersion} up to $z$. The change can also
be written more explicitly as 
\begin{align}
\tilde{A}_{x}(z,f) & =\sqrt{P_{0}\mathcal{G}(z)}\tilde{U}_{x}(z,f)\e^{\psi_{x}(f^{2})}\nonumber \\
\tilde{A}_{y}(z,f) & =\sqrt{P_{0}\mathcal{G}(z)}\tilde{U}_{y}(z,f)\e^{\psi_{y}(f^{2})}\label{eq:change_1_1}
\end{align}
where we defined
\begin{align*}
\psi_{x}(f^{2}) & =j\frac{C(z)(2\pi f)^{2}}{2}-j\phi_{x}\\
\psi_{y}(f^{2}) & =j\frac{C(z)(2\pi f)^{2}}{2}-j\phi_{y}
\end{align*}
 
\begin{align*}
\phi_{x}(z) & =\int_{0}^{z}\gamma(s)\mathcal{G}(s)\mbox{d} s\cdot(2P_{x}+P_{y})\\
\phi_{y}(z) & =\int_{0}^{z}\gamma(s)\mathcal{G}(s)\mbox{d} s\cdot(2P_{y}+P_{x})
\end{align*}

Differentiating Eq. (\ref{change_1}) gives
\begin{equation}
\frac{\partial\tilde{A}_{x}}{\partial z}=\sqrt{P_{0}\mathcal{G}(z)}\left(\frac{\partial\tilde{U}_{x}}{\partial z}+[\frac{g(z)-j\omega^{2}\beta_{2}(z)}{2}-j\underbrace{\gamma(z)\mathcal{G}(z)(2P_{x}+P_{y})}_{\phi_{X}'(z)}]\tilde{U}_{x}\right)\e^{j\psi_{x}(\omega^{2})}\label{eq:part1}
\end{equation}
and a dual equation holds for $\tilde{A}_{y}$. Substituting into
Eq. (\ref{eq:Sch_eq_f-1}) we get for the X component:
\begin{align*}
\sqrt{P_{0}\mathcal{G}(z)}\left(\frac{\partial\tilde{U}_{x}}{\partial z}+[\frac{g(z)-j\omega^{2}\beta_{2}(z)}{2}+j\gamma(z)\mathcal{G}(z)]\tilde{U}_{x}\right)\e^{\psi_{x}(f^{2})} & =\frac{g(z)-j\omega^{2}\beta_{2}(z)}{2}\tilde{U}_{x}(z,f)\sqrt{P_{0}\mathcal{G}(z)}\e^{\psi_{x}(f^{2})}\\
 & -j\gamma(z)\iint_{-\infty}^{\infty}\mbox{d} f_{1}\mbox{d} f_{2}P_{0}\mathcal{G}(z)\\
(\e^{\psi_{x}((f+f_{1}+f_{2})^{2})*}\tilde{U}_{x}^{*}(z,f+f_{1}+f_{2})\e^{\psi_{x}((f+f_{2})^{2}}\tilde{U}_{x}(z,f+f_{2}) & +\e^{\psi_{y}((f+f_{1}+f_{2})^{2})*}\tilde{U}_{y}^{*}\e^{\psi_{y}((f+f_{2})^{2})}\tilde{U}_{y})\\
\cdot & \sqrt{P_{0}\mathcal{G}(z)}\e^{\psi_{x}((f+f_{1})^{2})}\tilde{U}_{x}(z,f+f_{1}).
\end{align*}

Dividing both sides by $\sqrt{P_{0}\mathcal{G}(z)}e^{\psi_{x}(f^{2})}$
and simplifying, one gets:
\begin{align*}
\frac{\partial\tilde{U}_{x}}{\partial z}+j\gamma(z)\mathcal{G}(z)(2P_{x}+P_{y})\tilde{U}_{x} & =-j\gamma(z)\iint_{-\infty}^{\infty}\mbox{d} f_{1}\mbox{d} f_{2}P_{0}\mathcal{G}(z)\e^{-\psi_{x}(f^{2})}\\
(\e^{\psi_{x}((f+f_{1}+f_{2})^{2})*}\tilde{U}_{x}^{*}(z,f+f_{1}+f_{2})\e^{\psi_{x}((f+f_{2})^{2})}\tilde{U}_{x}(z,f+f_{2}) & +\e^{\psi_{y}((f+f_{1}+f_{2})^{2})*}\tilde{U}_{y}^{*}\e^{\psi_{y}((f+f_{2})^{2})}\tilde{U}_{y})\\
\cdot & \e^{\psi_{x}((f+f_{1})^{2}]}\tilde{U}_{x}(z,f+f_{1}).
\end{align*}

Now let's collect all the exponents of $\e^{(.)}$ together in the
r.h.s.. For the first term we have the exponent
\begin{align*}
-\psi_{x}(f^{2})+\psi_{x}((f+f_{1}+f_{2})^{2})*+\psi_{x}((f+f_{2})^{2})+\psi_{x}((f+f_{1})^{2}) & =\\
-[j\frac{C(z)\omega^{2}}{2}-j\phi_{x}]-[j\frac{C(z)(\omega+\omega_{1}+\omega_{2})^{2}}{2}-j\phi_{x}]+[j\frac{C(z)(\omega+\omega_{2})^{2}}{2}-j\phi_{x}]+[j\frac{C(z)(\omega+\omega_{1})^{2}}{2}-j\phi_{x}] & =\\
\frac{C(z)}{2}(-\omega^{2}-(\omega+\omega_{1}+\omega_{2})^{2}+(\omega+\omega_{2})^{2}+(\omega+\omega_{1})^{2}) & =\\
-C(z)\omega_{1}\omega_{2}
\end{align*}
where we used the relation
\begin{equation}
[\omega^{2}-(\omega+\omega_{1})^{2}-(\omega+\omega_{1})^{2}+(\omega+\omega_{1}+\omega_{2})^{2}]=2\omega_{1}\omega_{2}.\label{eq:lucky}
\end{equation}

For the second term we have the exponent
\begin{align*}
-\psi_{x}(f^{2})+\psi_{y}((f+f_{1}+f_{2})^{2})*+\psi_{y}((f+f_{2})^{2})+\psi_{x}((f+f_{1})^{2}) & =\\
-[j\frac{C(z)\omega^{2}}{2}-j\phi_{x}]-[j\frac{C(z)(\omega+\omega_{1}+\omega_{2})^{2}}{2}-j\phi_{y}]+[j\frac{C(z)(\omega+\omega_{2})^{2}}{2}-j\phi_{y}]+[j\frac{C(z)(\omega+\omega_{1})^{2}}{2}-j\phi_{x}] & =\\
\frac{C(z)}{2}(-\omega^{2}-(\omega+\omega_{1}+\omega_{2})^{2}+(\omega+\omega_{2})^{2}+(\omega+\omega_{1})^{2}) & =\\
-C(z)\omega_{1}\omega_{2}
\end{align*}

Hence finally we get the Manakov DMNLSE (M-DMNLSE) for X:
\begin{align}
\frac{\partial\tilde{U}_{x}(z,f)}{\partial z} & {\color{red}=+j\underbrace{\gamma(z)\mathcal{G}(z)(2P_{x}+P_{y})}_{\phi_{x}'(z)}\tilde{U}_{x}(z,f)}-j\gamma(z)\mathcal{G}(z)P_{0}\iint_{-\infty}^{\infty}\e^{-jC(z)\omega_{1}\omega_{2}}\nonumber \\
(\tilde{U}_{x}^{*}(z,f+f_{1}+f_{2})\tilde{U}_{x}(z,f+f_{2}) & +\tilde{U}_{y}^{*}(z,f+f_{1}+f_{2})\tilde{U}_{y}(z,f+f_{2}))\tilde{U}_{x}(z,f+f_{1})\mbox{d} f_{1}\mbox{d} f_{2}\label{eq:due}
\end{align}
and swapping indices x and y we get the dual equation for Y. In red
I have highlighted the main difference from the traditional result.

\subsection{Dual Polarization first-order Enhanced Regular Perturbation (DP-ERP1)}

The output field $\tilde{\boldsymbol{U}}(L,f)$ is obtained as the
following formal explicit solution of (\ref{eq:due}):
\begin{align}
\tilde{U}_{x}(L,f)-\tilde{U}_{x}(0,f) & =j\int_{0}^{L}\phi_{x}'(z)\tilde{U}_{x}(s,f)\mbox{d} s\nonumber \\
 & -jP_{0}\iint_{-\infty}^{\infty}\int_{0}^{L}\gamma(s)\mathcal{G}(s)\e^{-jC(s)\omega_{1}\omega_{2}}\cdot\nonumber \\
(\tilde{U}_{x}^{*}(s,f+f_{1}+f_{2})\tilde{U}_{x}(s,f+f_{2}) & +\tilde{U}_{y}^{*}(s,f+f_{1}+f_{2})\tilde{U}_{y}(s,f+f_{2}))\tilde{U}_{x}(s,f+f_{1})\mbox{d} s\mbox{d} f_{1}\mbox{d} f_{2}.\label{eq:DMNLSEsolx}
\end{align}

\selectlanguage{english}%
For systems where the spectrum of the propagating normalized signal
$\tilde{\boldsymbol{U}}(z,\omega)$ does not significantly vary along
$z$ we see that the exact solution (\ref{eq:DMNLSEsolx}) may be
approximated by the first-order regular perturbation (\foreignlanguage{american}{DP-ERP1})
solution, which we write for the X component:
\begin{align}
\tilde{U}_{x}(L,f)= & \tilde{U}_{x}(0,f)+j\underbrace{\phi_{x}(L)}_{(2P_{x}+P_{y})\mathcal{K}(0)}\tilde{U}_{x}(0,f)\nonumber \\
 & -jP_{0}\iint_{-\infty}^{\infty}\mathcal{K}(f_{1}f_{2})(\tilde{U}_{x}^{*}(0,f+f_{1}+f_{2})\tilde{U}_{x}(0,f+f_{2})+\nonumber \\
 & \tilde{U}_{y}^{*}(0,f+f_{1}+f_{2})\tilde{U}_{y}(0,f+f_{2}))\tilde{U}_{x}(0,f+f_{1})\mbox{d} f_{1}\mbox{d} f_{2}\label{eq:RP1x}
\end{align}
and a dual expression holds for Y by swapping indices $x\leftrightarrow y$,
where we defined 
\begin{equation}
\mathcal{K}(F)\triangleq\int_{0}^{L}\gamma(s)\mathcal{G}(s)\e^{-jC(s)(2\pi)^{2}F}ds\label{eq:Enne}
\end{equation}
as the un-normalized frequency kernel, a function of only the product
$F=f_{1}f_{2}$. We may also define the (cumulated) nonlinear phase
as 
\begin{equation}
\Phi_{NL}=P_{0}\mathcal{K}(0)\qquad[rad]\label{eq:phi_NL}
\end{equation}
and the normalized (scalar) frequency kernel, or briefly the \textbf{frequency
kernel }as 
\begin{equation}
\fmbox{\tilde{\eta}(F)=\frac{\mathcal{K}(F)}{\mathcal{K}(0)}=\frac{\int_{0}^{L}\gamma(s)\mathcal{G}(s)\e^{-jC(s)(2\pi)^{2}F}ds}{\int_{0}^{L}\gamma(s)\mathcal{G}(s)ds}}\label{eq:DMkernel-1}
\end{equation}
\foreignlanguage{american}{so that one finally gets a convenient form
of the RP1 perturbation as:
\[
\frac{\tilde{U}_{x,p}(L,f)}{-j\Phi_{NL}}=-A(f)+B(f)
\]
}with 
\[
A(f)\equiv\hat{P}_{Tx}\tilde{U}_{x}(0,f)
\]
 and 
\begin{align*}
B(f)\equiv & \iint_{-\infty}^{\infty}\mathcal{\tilde{\eta}}(f_{1}f_{2})(\tilde{U}_{x}^{*}(0,f+f_{1}+f_{2})\tilde{U}_{x}(0,f+f_{2})+\\
 & \tilde{U}_{y}^{*}(0,f+f_{1}+f_{2})\tilde{U}_{y}(0,f+f_{2}))\tilde{U}_{x}(0,f+f_{1})\mbox{d} f_{1}\mbox{d} f_{2}
\end{align*}
\foreignlanguage{american}{where $\hat{P}_{x}=P_{x}/P_{0}$, $\hat{P}_{y}=P_{y}/P_{0}$,
and $\hat{P}_{Tx}=(2\hat{P}_{x}+\hat{P}_{y})$.}

\selectlanguage{american}%
Now we must repeat the long calculations used to verify Joannisson's
result in \cite{arxiv}, and check if indeed the NLI power spectral
density obtained from the DP-ERP1 solution does not have the unwanted
``phase'' components, third line in (\ref{eq:pontus_X_sol}). We
need
\[
\frac{E[\tilde{U}_{x,p}(L,f)\tilde{U}_{x,p}^{*}(L,u)]}{\Phi_{NL}^{2}}=E[A(f)A^{*}(u)]-2\mbox{Re}(E[B(f)A^{*}(u)])+E[B(f)B^{*}(u)]
\]
 and in \cite{arxiv} we already calculated $E[B(f)B^{*}(u)]$, while
\begin{align*}
E[A(f)A^{*}(u)] & =\hat{P}_{Tx}^{2}E[\tilde{U}_{x}(0,f)\tilde{U}_{x}^{*}(0,u)]\\
 & =\hat{P}_{Tx}^{2}\hat{G}_{x}(f)\delta(f-u)
\end{align*}
where we used Theorem 1 in \cite{arxiv}. We finally have to evaluate
\begin{align*}
E[B(f)A^{*}(u)] & =\iint_{-\infty}^{\infty}\mbox{d} f_{1}\mbox{d} f_{2}\mathcal{\tilde{\eta}}(f_{1}f_{2})\cdot\\
 & \hat{P}_{Tx}\{E[\tilde{U}_{x}^{*}(0,f+f_{1}+f_{2})\tilde{U}_{x}(0,f+f_{2})\tilde{U}_{x}(0,f+f_{1})\tilde{U}_{x}^{*}(0,u)]+\\
 & E[\tilde{U}_{y}^{*}(0,f+f_{1}+f_{2})\tilde{U}_{y}(0,f+f_{2})\tilde{U}_{x}(0,f+f_{1})\tilde{U}_{x}^{*}(0,u)]\}.
\end{align*}

For 4 generic complex Gaussian RVs, theorem 2 in \cite{arxiv} specializes
to
\[
E[U_{1}^{*}U_{2}^{*}U_{3}U_{4}]=E[U_{1}^{*}U_{3}]E[U_{2}^{*}U_{4}]+E[U_{1}^{*}U_{4}]E[U_{2}^{*}U_{3}].
\]
Hence
\begin{align*}
E[\tilde{U}_{x}^{*}(0,f+f_{1}+f_{2})\tilde{U}_{x}(0,f+f_{2})\tilde{U}_{x}(0,f+f_{1})\tilde{U}_{x}^{*}(0,u)] & =\\
\underbrace{E[\tilde{U}_{x}^{*}(0,f+f_{1}+f_{2})\tilde{U}_{x}(0,f+f_{2})]}_{\hat{G}_{x}(f+f_{2})\delta(f_{1})}\underbrace{E[\tilde{U}_{x}^{*}(0,u)\tilde{U}_{x}(0,f+f_{1})]}_{\hat{G}_{x}(f+f_{1})\delta(u-f-f_{1})} & +\\
\underbrace{E[\tilde{U}_{x}^{*}(0,f+f_{1}+f_{2})\tilde{U}_{x}(0,f+f_{1})]}_{\hat{G}_{x}(f+f_{1})\delta(f_{2})}\underbrace{E[\tilde{U}_{x}^{*}(0,u)\tilde{U}_{x}(0,f+f_{2})]}_{\hat{G}_{x}(f+f_{2})\delta(u-f-f_{2})} & =\\
\hat{G}_{x}(f+f_{2})\hat{G}_{x}(f)\delta(f_{1})\delta(u-f)+\hat{G}_{x}(f+f_{1})\delta(f_{2})\hat{G}_{x}(f)\delta(u-f)
\end{align*}
and 
\begin{align*}
E[\tilde{U}_{y}^{*}(0,f+f_{1}+f_{2})\tilde{U}_{y}(0,f+f_{2})\tilde{U}_{x}(0,f+f_{1})\tilde{U}_{x}^{*}(0,u)] & =\\
\underbrace{E[\tilde{U}_{y}^{*}(0,f+f_{1}+f_{2})\tilde{U}_{y}(0,f+f_{2})]}_{\hat{G}_{y}(f+f_{2})\delta(f_{1})}\underbrace{E[\tilde{U}_{x}^{*}(0,u)\tilde{U}_{x}(0,f+f_{1})]}_{\hat{G}_{x}(f+f_{1})\delta(u-f-f_{1})} & +\\
\underbrace{E[\tilde{U}_{y}^{*}(0,f+f_{1}+f_{2})\tilde{U}_{x}(0,f+f_{1})]}_{0}\underbrace{E[\tilde{U}_{x}^{*}(0,u)\tilde{U}_{y}(0,f+f_{2})]}_{0} & =\\
\hat{G}_{y}(f+f_{2})\hat{G}_{x}(f)\delta(f_{1})\delta(u-f) & .
\end{align*}
Thus
\begin{align*}
E[B(f)A^{*}(u)]= & \delta(u-f)\iint_{-\infty}^{\infty}\mathcal{\tilde{\eta}}(f_{1}f_{2})\hat{P}_{Tx}[\hat{G}_{x}(f+f_{2})\hat{G}_{x}(f)\delta(f_{1})+\\
+ & \hat{G}_{x}(f+f_{1})\delta(f_{2})\hat{G}_{x}(f)+\hat{G}_{y}(f+f_{2})\hat{G}_{x}(f)\delta(f_{1})]\mbox{d} f_{1}\mbox{d} f_{2}\\
= & \delta(u-f)\hat{G}_{x}(f)\hat{P}_{Tx}\mathcal{\tilde{\eta}}(0)(2\int_{-\infty}^{\infty}\hat{G}_{x}(f_{2})\mbox{d} f_{2}+\int_{-\infty}^{\infty}\hat{G}_{y}(f_{2})\mbox{d} f_{2})\\
= & \delta(u-f)\hat{G}_{x}(f)\hat{P}_{Tx}(2\hat{P}_{x}+\hat{P}_{y}).
\end{align*}

Finally from \cite{arxiv} we already know that
\begin{align*}
E[B(f)B^{*}(u)] & =\delta(u-f)\{2\iintop_{-\infty}^{\infty}|\tilde{\eta}(f_{1}f_{2})|^{2}\hat{G}_{x}(f+f_{1}+f_{2})\hat{G}_{x}(f+f_{1})\hat{G}_{x}(f+f_{2})\mbox{d} f_{1}\mbox{d} f_{2}\\
 & +\iintop_{-\infty}^{\infty}|\tilde{\eta}(f_{1}f_{2})|^{2}\hat{G}_{y}(f+f_{1}+f_{2})\hat{G}_{x}(f+f_{1})\hat{G}_{y}(f+f_{2})\mbox{d} f_{1}\mbox{d} f_{2}\\
 & +\hat{G}_{x}(f)(2\hat{P}_{x}+\hat{P}_{y})^{2}\}.
\end{align*}

Therefore, getting rid of $\delta(u-f)$ we have
\begin{align*}
\frac{\hat{G}_{x,p}(f)}{\Phi_{NL}^{2}} & {\color{red}=\{\hat{P}_{Tx}^{2}\hat{G}_{x}(f)}\\
 & {\color{red}-2\hat{G}_{x}(f)\hat{P}_{Tx}(2\hat{P}_{x}+\hat{P}_{y})}\\
 & {\color{red}+\hat{G}_{x}(f)(2\hat{P}_{x}+\hat{P}_{y})^{2}\}}\\
 & +2\iintop_{-\infty}^{\infty}|\tilde{\eta}(f_{1}f_{2})|^{2}\hat{G}_{x}(f+f_{1}+f_{2})\hat{G}_{x}(f+f_{1})\hat{G}_{x}(f+f_{2})\mbox{d} f_{1}\mbox{d} f_{2}\\
 & +\iintop_{-\infty}^{\infty}|\tilde{\eta}(f_{1}f_{2})|^{2}\hat{G}_{y}(f+f_{1}+f_{2})\hat{G}_{x}(f+f_{1})\hat{G}_{y}(f+f_{2})\mbox{d} f_{1}\mbox{d} f_{2}
\end{align*}
and we recognize that the red curly bracket is zero since by definition
$\hat{P}_{Tx}\equiv(2\hat{P}_{x}+\hat{P}_{y}).$ Hence indeed the
DP-ERP ansatz removes the unwanted ``phase'' terms in the NLI PSD. 

\section{Conclusions}

This report is conceptually very important, as it explains the following
apparent contradiction about the RP1 solution of the DMNLSE:

It is known from \cite{RP} that the RP1 solution largely overestimates
the true DMNLSE solution, and holds only at extremely small nonlinear
phases. How can it then accurately reproduce the nonlinear interference
(NLI) power spectral density, as claimed by Turin's group \cite{carena_JLT}?

The explanation is the following: if instead of the RP1 solution we
use the ERP1 solution \cite{RP}, and as extended here to dual polarization
and called the \textbf{DP-ERP1} solution, then the numerical predictions
are very close to the true field, and thus also the NLI variance predictions
are close to real. Moreover, the DP-ERP1 predicted NLI PSD is exactly
that reported by Turin's group \cite{carena_JLT}, without the puzzling
``phase term'' in the RP1 result in \cite{Pontus}.

\section*{Appendix 1}

In this Appendix we prove Theorem 3 in the text. Assuming jointly
stationary circular complex Gaussian zero-mean processes $A(t),B(t),C(t),D(t),E(t),F(t)$,
we have by using (\ref{eq:papoulis}) in (\ref{eq:iss_6}):
\begin{align*}
T\triangleq E\left[\tilde{A}(f+f_{1})\tilde{B}^{*}(f+f_{1}+f_{2})\tilde{C}(f+f_{2})\tilde{D}^{*}(u+f_{3})\tilde{E}(u+f_{3}+f_{4})\tilde{F}^{*}(u+f_{4})\right] & =\\
\underbrace{E[\tilde{B}^{*}(f+f_{1}+f_{2})\tilde{A}(f+f_{1})]}_{G_{ab}(f+f_{1})\delta(f_{2})}\underbrace{E[\tilde{D}^{*}(u+f_{3})\tilde{C}(f+f_{2})]}_{G_{cd}(f+f_{2})\delta(u+f_{3}-f-f_{2})}\underbrace{E[\tilde{F}^{*}(u+f_{4})\tilde{E}(u+f_{3}+f_{4})]}_{G_{ef}(u+f_{3}+f_{4})\delta(f_{3})} & +\\
\underbrace{E[\tilde{B}^{*}(f+f_{1}+f_{2})\tilde{A}(f+f_{1})]}_{G_{ab}(f+f_{1})\delta(f_{2})}\underbrace{E[\tilde{D}^{*}(u+f_{3})\tilde{E}(u+f_{3}+f_{4})]}_{G_{ed}(u+f_{3}+f_{4})\delta(f_{4})}\underbrace{E[\tilde{F}^{*}(u+f_{4})\tilde{C}(f+f_{2})]}_{G_{cf}(f+f_{2})\delta(u+f_{4}-f-f_{2})} & +\\
\underbrace{E[\tilde{B}^{*}(f+f_{1}+f_{2})\tilde{C}(f+f_{2})]}_{G_{cb}(f+f_{2})\delta(f_{1})}\underbrace{E[\tilde{D}^{*}(u+f_{3})\tilde{A}(f+f_{1})]}_{G_{ad}(f+f_{1})\delta(u+f_{3}-f-f_{1})}\underbrace{E[\tilde{F}^{*}(u+f_{4})\tilde{E}(u+f_{3}+f_{4})]}_{G_{ef}(u+f_{3}+f_{4})\delta(f_{3})} & +\\
\underbrace{E[\tilde{B}^{*}(f+f_{1}+f_{2})\tilde{C}(f+f_{2})]}_{G_{cb}(f+f_{2})\delta(f_{1})}\underbrace{E[\tilde{D}^{*}(u+f_{3})\tilde{E}(u+f_{3}+f_{4})]}_{G_{ed}(u+f_{3}+f_{4})\delta(f_{4})}\underbrace{E[\tilde{F}^{*}(u+f_{4})\tilde{A}(f+f_{1})]}_{G_{af}(f+f_{1})\delta(u+f_{4}-f-f_{1})} & +\\
\underbrace{E[\tilde{B}^{*}(f+f_{1}+f_{2})\tilde{E}(u+f_{3}+f_{4})]}_{G_{eb}(u+f_{3}+f_{4})\delta(f+f_{1}+f_{2}-u-f_{3}-f_{4})}\underbrace{E[\tilde{D}^{*}(u+f_{3})\tilde{A}(f+f_{1})]}_{G_{ad}(f+f_{1})\delta(u+f_{3}-f-f_{1})}\underbrace{E[\tilde{F}^{*}(u+f_{4})\tilde{C}(f+f_{2})]}_{G_{cf}(f+f_{2})\delta(u+f_{4}-f-f_{2})} & +\\
\underbrace{E[\tilde{B}^{*}(f+f_{1}+f_{2})\tilde{E}(u+f_{3}+f_{4})]}_{G_{eb}(u+f_{3}+f_{4})\delta(f+f_{1}+f_{2}-u-f_{3}-f_{4})}\underbrace{E[\tilde{D}^{*}(u+f_{3})\tilde{C}(f+f_{2})]}_{G_{cd}(f+f_{2})\delta(u+f_{3}-f-f_{2})}\underbrace{E[\tilde{F}^{*}(u+f_{4})\tilde{A}(f+f_{1})]}_{G_{af}(f+f_{1})\delta(u+f_{4}-f-f_{1})} & =
\end{align*}
thus
\begin{align*}
T & =G_{ab}(f+f_{1})G_{cd}(f+f_{2})G_{ef}(u+f_{3}+f_{4})\delta(f_{2})\delta(f_{3})\delta(u+f_{3}-f-f_{2})\\
 & +G_{ab}(f+f_{1})G_{ed}(u+f_{3}+f_{4})G_{cf}(f+f_{2})\delta(f_{2})\delta(f_{4})\delta(u+f_{4}-f-f_{2})\\
 & +G_{cb}(f+f_{2})G_{ad}(f+f_{1})G_{ef}(u+f_{3}+f_{4})\delta(f_{1})\delta(f_{3})\delta(u+f_{3}-f-f_{1})\\
 & +G_{cb}(f+f_{2})G_{ed}(u+f_{3}+f_{4})G_{af}(f+f_{1})\delta(f_{1})\delta(f_{4})\delta(u+f_{4}-f-f_{1})\\
 & +G_{eb}(u+f_{3}+f_{4})G_{ad}(f+f_{1})G_{cf}(f+f_{2})\cdot\\
 & \cdot\delta(u+f_{4}-f-f_{2})\delta(u+f_{3}-f-f_{1})\delta(f+f_{1}+f_{2}-u-f_{3}-f_{4})\\
 & +G_{eb}(u+f_{3}+f_{4})G_{cd}(f+f_{2})G_{af}(f+f_{1})\cdot\\
 & \cdot\delta(u+f_{4}-f-f_{1})\delta(u+f_{3}-f-f_{2})\delta(f+f_{1}+f_{2}-u-f_{3}-f_{4}).
\end{align*}

Now we use the sampling property of the delta to write, e.g. for the
first line where $f_{2}=0$ and $f_{3}=0$, 
\[
G_{ab}(f+f_{1})G_{cd}(f)G_{ef}(u+f_{4})\delta(f_{2})\delta(f_{3})\delta(u-f)
\]
and e.g. for the last line where $u+f_{4}=f+f_{1}$ and $u+f_{3}=f+f_{2}$
which we add up to get
\[
{\color{green}u+f_{3}+f_{4}=(f-u)+f+f_{1}+f_{2}}
\]
whence
\[
{\color{red}f+f_{1}+f_{2}-u-f_{3}-f_{4}}={\color{red}u-f}
\]
so that the last line writes as
\begin{align*}
 & G_{eb}({\color{green}u+f_{3}+f_{4}})G_{cd}(f+f_{2})G_{af}(f+f_{1})\delta(u+f_{4}-f-f_{1})\delta(u+f_{3}-f-f_{2})\delta({\color{red}f+f_{1}+f_{2}-u-f_{3}-f_{4}})=\\
 & G_{eb}({\color{green}(f-u)+f+f_{1}+f_{2}})G_{cd}(f+f_{2})G_{af}(f+f_{1})\cdot\delta(u+f_{4}-f-f_{1})\delta(u+f_{3}-f-f_{2})\delta({\color{red}u-f})\stackrel{(\mbox{use\,}u=f)}{=}\\
 & G_{eb}({\color{green}f+f_{1}+f_{2}})G_{cd}(f+f_{2})G_{af}(f+f_{1})\delta(f_{4}-f_{1})\delta(f_{3}-f_{2})\delta({\color{red}u-f}).
\end{align*}

We therefore get
\begin{align*}
T & =G_{ab}(f+f_{1})G_{cd}(f)G_{ef}(f+f_{4})\delta(f_{2})\delta(f_{3})\delta(u-f)\\
 & +G_{ab}(f+f_{1})G_{ed}(f+f_{3})G_{cf}(f)\delta(f_{2})\delta(f_{4})\delta(u-f)\\
 & +G_{cb}(f+f_{2})G_{ad}(f)G_{ef}(f+f_{4})\delta(f_{1})\delta(f_{3})\delta(u-f)\\
 & +G_{cb}(f+f_{2})G_{ed}(f+f_{3})G_{af}(f)\delta(f_{1})\delta(f_{4})\delta(u-f)\\
 & +G_{eb}(f+f_{1}+f_{2})G_{ad}(f+f_{1})G_{cf}(f+f_{2})\delta(f_{4}-f_{2})\delta(f_{3}-f_{1})\delta(u-f)\\
 & +G_{eb}(f+f_{1}+f_{2})G_{cd}(f+f_{2})G_{af}(f+f_{1})\delta(f_{4}-f_{1})\delta(f_{3}-f_{2})\delta(u-f).
\end{align*}
whence the final form (\ref{eq:form_compact-1}) given in Theorem
3.
\end{document}